\documentclass[12pt]{article}
\usepackage[left=1.5in,right=1.5in,top=1.2in,bottom=1in]{geometry}
\usepackage[utf8]{inputenc}
\usepackage[sectionbib]{natbib}
\usepackage{array,epsfig,fancyheadings,rotating}
\usepackage[]{hyperref}  
\usepackage{sectsty, secdot}
\usepackage{amsmath}
\usepackage{amssymb}
\usepackage{amsfonts}
\usepackage{multirow}
\usepackage{amsthm}

\usepackage[mathscr]{eucal}
\usepackage{hyperref}
\usepackage{dlfltxbcodetips}
\usepackage{bbm} 
\usepackage{mathtools}
\usepackage{amsfonts}
\usepackage{amsmath} 
\usepackage{amssymb}
\usepackage{fancybox} 
\usepackage{graphicx}
\usepackage{bm}
\usepackage{latexsym}
\usepackage{tcolorbox,enumitem,setspace}

\usepackage{caption}
\usepackage{subcaption}

\newcommand{\diag}{\text{diag}}
\newcommand{\obs}{\text{obs}}

\newcommand{\mM}{\mathscr{M}}

\newcommand{\mP}{\mathscr{P}}

\newcommand{\mbP}{\mathbb{P}}

\newcommand{\mbR}{\mathbb{R}}

\newcommand{\bd}{\bm{d}}

\newcommand{\bx}{\bm{x}}

\newcommand{\bz}{\bm{z}}

\newcommand{\bA}{\bm{A}}
\newcommand{\bB}{\bm{B}}

\newcommand{\bD}{\bm{D}}

\newcommand{\bI}{\bm{I}}

\newcommand{\bL}{\bm{L}}

\newcommand{\bP}{\bm{P}}

\newcommand{\bU}{\bm{U}}

\newcommand{\bX}{\bm{X}}

\newcommand{\bZ}{\bm{Z}}

\newcommand{\blambda}{\mbox{\boldmath$\lambda$}}

\newcommand{\btheta}{\boldsymbol{\theta}}

\newcommand{\one}{\mathbbm{1}}

\newcommand{\dsum}{\displaystyle\sum\limits}

\newcommand{\dprod}{\displaystyle\prod\limits}

\newcommand{\norm}[1]{\lVert#1\rVert}

\newcommand{\btt}{\begin{box}}
\newcommand{\ett}{\end{box}}

\newcommand{\btheorem}{\begin{bclogo}[couleur={rgb:orange,0;yellow,0;white,1},arrondi=0.1,logo=\bcplume,ombre=true]{Theorem}}
\newcommand{\ettheorem}{\end{bclogo}}

\newcommand{\bst}{\begin{bclogo}[couleur={rgb:orange,1;yellow,1;white,0.5},arrondi=0.1,logo=\bcpanchant]}
\newcommand{\est}{\end{bclogo}}

\newcommand{\ba}{\begin{array}{llllllllll}}
\newcommand{\ea}{\end{array}}

\newcommand{\bea}{\begin{equation}\begin{array}{llllllllll}}
\newcommand{\eea}{\end{array}\end{equation}}
\newcommand{\be}{\begin{equation}\begin{array}{lllllllllllllllll}}
\newcommand{\beno}{\begin{equation}\begin{array}{lllllllllllll}\nonumber}
\newcommand{\ee}{\end{array}\end{equation}}
\newcommand{\bel}{\begin{equation}\begin{array}{lllllllllllll}\nonumber}
\newcommand{\eel}{\Box\end{array}\end{equation}}
\newcommand{\bi}{\begin{itemize}}
\newcommand{\ei}{\end{itemize}}
\newcommand{\ben}{\begin{enumerate}}
\newcommand{\een}{\end{enumerate}}
\newcommand{\bbq}{\begin{quote}\bf\em}
\newcommand{\ebq}{\end{quote}}

\renewcommand{\=}{&=&}

\newcommand{\hide}[1]{}
\newcommand{\ghost}[1]{}

\newcommand{\bcom}{\begin{comment}\em}
\newcommand{\ecom}{\end{comment}}
\newcounter{counterexample}

\newcounter{ttproof}
\newcounter{pproof}
\newcounter{ccproof}
\newcounter{llproof}
\newcounter{com}

\newcounter{assumption}

\hypersetup{
    colorlinks,
    linkcolor={red!50!black},
    citecolor={blue!50!black},
    urlcolor={blue!80!black}
}

\begin{document}

\fontsize{12}{14pt plus.8pt minus .6pt}\selectfont \vspace{0.8pc}
\centerline{\large\bf MODEL SELECTION FOR NETWORK DATA }
\vspace{2pt} 
\centerline{\large\bf BASED ON SPECTRAL INFORMATION}
\vspace{.4cm} 
\centerline{Jairo Ivan Pe\~{n}a Hidalgo AND Jonathan R. Stewart} 
\vspace{.4cm} 
\centerline{\it Florida State University}
 \vspace{.55cm} \fontsize{9}{11.5pt plus.8pt minus.6pt}\selectfont

\begin{quotation}
\noindent {\it Abstract:}
We introduce a new methodology for model selection in the context of modeling network data. 
The statistical network analysis literature has developed many different classes of network data models, 
with notable model classes including stochastic block models, 
latent position models, 
and exponential families of random graph models. 
A persistent question in the statistical network analysis literature lies in understanding how to 
compare different models for the purpose of model selection and evaluating goodness-of-fit,
especially when models have different mathematical foundations. 
In this work, we develop a novel non-parametric method for model selection in network data settings 
 which exploits the information contained in the spectrum of the graph Laplacian 
in order to obtain  
a measure of goodness-of-fit for a defined set of network data models. 
We explore the performance of our proposed methodology to popular classes of network data models through numerous simulation studies,
demonstrating the 
practical utility of our method through two applications.  

\vspace{9pt}
\par
\end{quotation}\par

\section{Introduction}
\label{sec:intro}

Network data have witnessed a surge of interest across a variety of fields and disciplines in recent decades,
including the study of
social networks \citep{ergm.book},
network epidemiology (involving the spread of disease through networks of contacts) \citep{morris2004},
covert networks of criminal activity and terrorism \citep{CoDiToDaKo20},
brain networks \citep{ObFa17},
financial markets \citep{FiLu17},
and more.
Network data, as a data structure,
are typically represented as a graph \citep{Ko09a},
consisting  
of a set of nodes representing the elements of a population of interest (e.g., researchers in a collaboration network) 
and a set of pairwise observations or measurements between nodes 
represented as edges between nodes (e.g., co-authorship on a paper). 
\hide{
An important problem in statistical analysis of network data is how to select a model for an observed network
in settings where the models may not share the same mathematical foundations
(e.g., parametric Markov random field specifications of random graphs versus latent space models for network data).
Much of the developed methodology and theory for model selection in network data settings 
has been developed for 
particular classes of models, 
and there is a need for a more comprehensive methodology for model selection in network data settings. 
We expand the literature by introducing a novel  
non-parametric methodology for model selection
that can be applied to a broad class of models under weak assumptions by 
utilizing information in the spectrum of the graph Laplacian. 
}

Many classes of models have been proposed and developed to study and model network data. 
A non-exhaustive review of some of the more prominent examples include 
exponential families of random graph models (ERGMs) \citep[e.g.,][]{ergm.book, ScKrBu17},
stochastic block models (SBMs) \citep[e.g.,][]{HoLaLe83,AnWaFa92,wang2017likelihood},
latent position models (LPMs) \citep[e.g.,][]{HoRaHa02,sewell2015latent,TaSuPr13,dot-product-graph},
and more. 
Each class offers a unique mathematical platform for constructing models of networks from observed network data,
with respective strengths and weaknesses. 
The exponential family class provides a flexible parametric platform for building models of networks with dependent edges. 
In contrast, 
stochastic block models can capture network structure and clustering of nodes through a discrete latent space, 
whereas latent position models capture network structure and edge dependence through latent node positions in (e.g.) 
a latent Euclidean space. 

A persistent challenge in statistical network analysis applications is how to compare
different models and select models for specific network data sets. 
At present, 
the literature has primarily focused on model selection problems within each class of models,
tailoring methods to specific classes of models 
(SBMs: \citet{wang2017likelihood,latouche2014model}; 
ERGMs: \citet{HuGoHa08}; LSMs: \citet{ryan2017bayesian}).
As a result,
there is a gap in the literature which explores methods for  
comparing model fit or performing model selection across models from different mathematical platforms, 
e.g., 
comparing an ERGM to an SBM to an LPM. 
In this work, 
we introduce a novel
non-parametric methodology for model selection in network data settings 
that can be applied to a broad class of models under weak assumptions,
capable of facilitating comparison of models with different mathematical foundations. 
Our method utilizes information 
in the spectrum of the graph Laplacian in order to select a best fitting model for an observed network,
and essentially only requires the ability to simulate adjacency matrices 
from candidate models and compute eigenvalues of 
the graph Laplacian derived from the adjacency matrices. 
%

The rest of the paper is organized as follows. 
Section \ref{sec:spectral} reviews spectral properties of the graph Laplacian for networks and 
motivates the use of spectral information in the model selection problem for network data.  
Our proposed methodology is introduced in Section \ref{sec:methods}. 
We present experimental studies and simulations in Section \ref{sec:sim}, 
and two applications of our methodology in Section \ref{sec:applications}. 

\section{Spectral properties of the graph Laplacian} 
\label{sec:spectral}

We consider simple undirected networks defined on a set of $N$ nodes with corresponding adjacency matrix 
$\bX \in \{0, 1\}^{N \times N}$, where $X_{i,j}=1$ corresponds to the event that there is an  
edge between nodes $i$ and $j$ and $X_{i,j} = 0$ otherwise. 
We adopt the standard conventions that $X_{i,j} = X_{j,i}$ an $X_{i,i} = 0$. 
Extensions of our methodology to directed networks is discussed in Section \ref{sec:sim}.
Extensions to networks with valued edges is possible, 
but beyond the scope of this work. 
Let $\bd =\text{deg}(\bX) =(\sum_{j=1}^N X_{i,j} : i = 1, \ldots, N)  \in \mathbb{R}^N$ be the vector of node degrees of the network. The Laplacian matrix, also called the graph Laplacian, is defined as $\bL(\bX) \coloneqq \diag(\bd) - \bX$, where $\diag(\bd)$ is the $N \times N$  diagonal matrix with diagonal $\bd$. 
Since $\bL(\bX)$ is symmetric and positive
semi-definite \citep{BaHw12},  the eigenvalues of $\bL(\bX)$ will all be real and non-negative. 
Throughout, 
we will let $\blambda \in \mathbb{R}^N$ 
denote the vector of ordered eigenvalues (from smallest to largest) of the Laplacian matrix $\bL(\bX)$. 
The vector $\blambda$ will depend on the adjacency matrix $\bX$ through $\bL(\bX)$, 
however, 
for ease of presentation, 
we do not make this dependence explicit notationally,
as it will be clear contextually.  

Eigenvalues of Laplacian matrices encode many well known properties of a network. 
For example, the multiplicity of the eigenvalue $0$ corresponds to the number of connected components in the network \citep{BaHw12}. 
The second smallest eigenvalue (possibly $0$) is known as the algebraic connectivity \citep{Fm73}, 
and measures the overall connectivity of a graph \citep{Mn07}. It is also used in stablishing Cheeger inequalities \citep{DlNfmM06}, which have applications in image segmentation \citep{SjMj00}, graph clustering \citep{CtCl13} and expander graphs \citep{HhLnWa06}.
 The subsequent eigenvalues of the Laplacian matrix are related to the minimal cuts (weighted edge deleting) required to partition a network \citep{BollobasNikiforov04}. 

Two undirected graphs with adjacency matrices $\bA$ and $\bB$ are isomorphic if there exists a permutation matrix
$\bP$ such that $\bA = \bP \bB \bP^t$,
which requires that the adjacency matrices be similar $\bA = \bP \, \bB \bP^{-1}$,
noting that a permutation matrix $\bP$ satisfies $\bP^{t} = \bP^{-1}$.
In such cases, 
the corresponding graph Laplacian matrices will be similar as well:  
\beno
L(\bA) 
\,=\, \deg(\bP \bB \bP^t) - \bP \, \bB \, \bP^t  
\,=\, \bP \deg(\bB) \bP^t - \bP \, \bB \, \bP^t  
\,=\, \bP \, L(\bB) \, \bP^t.
\ee 
Consequently,
since $L(\bB)$ is Hermitian,
there exists an eigen decomposition $L(\bB) = \bU \bD \, \bU^t$. 
Hence, 
$L(\bA) = \bP \, (\bU \bD \, \bU^t) \, \bP^t = (\bP \, \bU) \, \bD \, (\bP \, \bU)^t$.
As a result, if $\blambda$ is a vector of eigenvalues of $L(\bB)$,
it is also a vector of eigenvalues of $L(\bA)$. 
In our context, 
this implies one can always differentiate two non-isomorphic networks if their eigenvalues are different.
The reverse result is not true in general. 
There are graphs possessing the same eigenvalue decomposition 
(referred to as cospectral or isospectral) which are not isomorphic \citep{Cve80}. 
However, 
numerical evidence suggests that the fraction of (non-isomorphic) cospectral graphs tends to zero as the number of nodes in a graph grows 
\citep{BaHw12}.

Several applications of spectral decomposition of the Laplacian matrix have been proposed in the network analysis literature. 
For example, spectral clustering \citep{von07} is a well known clustering algorithm based on the leading eigenvectors of the Laplacian of a similarity matrix. 
\cite{LeiRa15} established, 
under mild conditions, 
the consistency of the spectral clustering method for stochastic block models. 
Another example is in \cite{Ne06},
where a family of community detection algorithms were proposed for networks based on the spectral decomposition 
of the graph Laplacian. 
Lastly, 
\citet{SjLb15} proposed a statistic for evaluating goodness-of-fit for network models 
reminiscent of the $R^2$ statistic in regression settings, 
which compares eigenvalues of the graph Laplacian generated from a model fit
to the eigenvalues of the graph Laplacian from a pre-specified {\it null} model 
(typically taken to be a Bernoulli random graph model, referred to as a density only model).

In light of these results, it is natural to regard the vector of eigenvalues $\blambda$ as a signature of a network, containing important topographical and structural information which can be exploited 
for the purposes of model selection. 
Our proposed methodology compares the empirical distribution of the spectrum of the graph Laplacian of 
candidate models to that of an observed network.
Our methodology is motivated by the following considerations regarding properties of the  graph Laplacian.  
 
First, if the true data generating process is in the list of candidate models, 
the observed eigenvalues derived from an observed network  
are expected to fall within the spectral distribution of the data generating process. 
If, in practice, none of the proposed models are the true generating process, 
candidate models can still be assessed by their ability to capture the spectrum of the observed graph Laplacian,
providing a means for developing a methods for model selection.
Second,
we can obtain a relative measure of fit among competing models depending on how well the 
spectrum of the observed graph Laplacian is captured by candidate models,
providing a means to not only select a best fitting model, 
but also to compare the fit of the best fitting model to unselected alternatives.  
Third, 
our methodology requires no parametric assumptions on the data generating process 
and is able to compare models across different mathematical platforms,
including models which do not have a well-defined likelihood function 
or which are constructed through a stochastic process, 
an example of which are agent-based models \citep[e.g.,][]{Snijders10, Jackson22} or generative algorithms based on preferential attachment models \citep[e.g.,][]{Barabasi99, Zeng13}.

\section{Methodology}
\label{sec:methods}

\begin{table}
\begin{center}
\begin{tcolorbox}[width=\linewidth]
{\bf Model selection procedure:} 
\begin{enumerate}[noitemsep]
\item Simulate $K$  networks $\bX^{(m,1)}, \ldots, \bX^{(m,K)}$ from each of the candidate models model $\mM_m \in \{\mM_1, \ldots \mM_M\}$.  
\item For each $\bX^{(m,k)}$,  
compute the Laplacian matrix $\bL(\bX^{(m,k)})$ and the corresponding vector of eigenvalues $\blambda^{(m,k)} \in \mbR^{N}$. 
\item Construct a design matrix $\bD \in \mbR^{(K M)\times N}$
by stacking the $KM$ vectors of eigenvalues $\blambda^{(m,k)}$ to form the rows of $\bD$. 
\item Train a classifier $\mP : \mbR^{N} \mapsto \{1, \ldots, M\}$ 
to predict a  model  $m^\star \in \{1, \ldots, M\}$ 
using the $K$ simulated vectors of eigenvectors  $\blambda^{(m,k)}$ for each class $m \in \{1, \ldots, M\}$ 
contained in the design matrix $\bD$. 
Feature engineering is advised at this stage. 
\item Compute the Laplacian matrix $\bL(\bX_{\obs})$ for the observed network $\bX_{\obs}$ 
and the corresponding vector of eigenvalues $\blambda_{\obs}$. 
\item Predict a class $m^\star = \mP(\blambda_{\obs})$ for the observed network using the trained classifier 
from Step 4 and set $\mM^\star = \mM_{m^\star}$. 
\end{enumerate}
\end{tcolorbox}
\end{center}
\caption{\label{tab:algorithm} Description of the model selection algorithm.} 
\end{table}

We outline a methodology for model selection in network data settings which exploits the spectral properties 
of the graph Laplacian, 
motivated by considerations in the previous section.
We assume throughout that the  network is completely observed,
denoted by $\bX_{\obs}$. 
The corresponding observed vector of eigenvalues of the
observed graph Laplacian $\bL(\bX_{\obs})$ is denoted by $\blambda_{\obs}$.
Our fundamental inferential goal is to select a best fitting model for the observed network $\bX_{\obs}$
from a set of candidate models $\{\mM_1, \ldots, \mM_M\}$ ($M \geq 2$).
We frame the problem as a classification problem and aim to construct
a classifier $\mP : \mbR^{N} \mapsto \{1, \ldots, M\}$ trained on the spectrum of the graph Laplacian 
for each of the candidate models in order to predict
a class $m^\star \in \{1, \ldots, M\}$ for a given vector of eigenvalues,
namely $\blambda_{\obs}$.
We present our model selection method algorithm in Table \ref{tab:algorithm}.

\subsection{Selection of classifier}

Real life networks can possess hundreds, thousands or even millions of nodes.
As the dimension of the vector of eigenvalues of the graph Laplacian matrices is equal to the number of nodes in the network,  
classification methods based on eigenvalues of the Laplacian matrix will be prone to the usual pitfalls 
of high dimensional classification problems. 
The literature for classification methods is quite extensive, 
which makes the choice of classifier a critical step in our methodology, 
although we show in Section \ref{sec:sim} that the effect of the choice of classifier 
may not have a significant effect on the results of our methodology under certain circumstances. 
In light of these results, 
we consider practical concerns of the implementation of the choice of classifier.

Linear discriminant analysis, 
 which requires the computation of the inverse of a covariance matrix,
has been shown in practice to suffer a decay in performance as the number of variables increases and the sample size is fixed \citep{BpEl04}. 
Alternative methods include support vector machines, neural networks, random forests, and boosting algorithms, which 
generally perform well in high-dimensional settings \citep{HaTiFr11}. 
Within this class is the eXtreme Gradient Boosting (XGBoost) method,
which offers both scalability and  state-of-the-art performance \citep{Chen16}. 
In the rest of this paper we use exclusively XGBoost, 
with the notable exception being Simulation study 5 in Section \ref{sec:sim},
in which we compare the performance of different classifiers to establish the claim made earlier in this section.

\subsection{Relative measure of goodness-of-fit}

Many classification algorithms return more than just a predicted class, 
often returning a vector of propensity scores $\bm{s} = (s_1, \ldots, s_M)$
with the property that $\norm{\bm{s}}_1 = 1$. 
If several models were considered, the propensity scores for many of the models can shrink 
simply because of the larger number of classes being considered,
meaning that the interpretation of propensity scores $s_1, \ldots, s_M$ can depend on $M$. 
To overcome this issue and facilitate the comparison of fit between models,
we propose to normalize the propensity scores to obtain a measure of goodness-of-fit which 
is independent of the number of candidate models $M$.  
To this end,
we define 
\beno
\tilde{s}_i 
\= \dfrac{s_i}{\norm{\bm{s}}_{\infty}},
&& i = 1, \ldots, M,
\ee
to be the normalized score, which is equal to $1$ for the highest scoring model. 
For all remaining models,
the normalized score is a measure of the fit of the model  relative to the highest scoring model. 
By rescaling all propensity scores in this manner, 
the number of models $M$ which is considered in the candidate set of models has no effect on the interpretation of the (relative) propensity scores.

\hide{
As an example, if we were to compare three models with propensity scores $(0.08,0.16,0.76)$, the relative scores would be $(0.105,0.210,1)$. We interpret this to mean that the first two models account for less than $21\%$ of the evidence present in the best performing model. We emphasize that, by rescaling all propensity scores, the number of models to be considered has no effect on the interpretation of the (relative) propensity scores.
}

\section{Simulation studies}
\label{sec:sim}

We conduct a number of simulation studies to demonstrate the potential of our proposed methodology. 
Specifically, 
we aim to examine the extent to which the signature of a network 
is contained within the spectrum of the graph Laplacian. 
Simulation studies permit knowledge of the true data-generating model, 
which facilitates empirical studies which aim to clarify the conditions under which our proposed methodology 
is able to successfully differentiate different network models and structural properties of networks.  

\hide{
In this section,  we present three simulation studies which demonstrate the potential of our methodology.
The first explores the interdependence of effect size and network size in  
exponential-family random graph models \citep{ergm.book}. 
The second demonstrates our methodology can be extended to direct networks 
and is able to detect important network features unique to directed networks, 
such as reciprocity. 
The third explores the potential of our methodology to correctly select 
the dimension of latent spaces in latent position models. 
}

\subsection{Simulation study 1: curved exponential families} 

We study the performance of our methodology on curved exponential families, 
which have gained popularity in the social network analysis community \citep[e.g.,][]{SnPaRoHa04,HuHa06}, 
as well as other applications \citep[e.g.,][]{ObFa17,ScKrBu17,StLo21}. 
The prominence of curved exponential family parameterizations for random graph models emerged out of a desire to solve 
challenges related to degeneracy and fitting of early and ill-posed model specifications \citep{SnPaRoHa04}. 
Additionally, 
curved exponential family parameterizations are able to parsimoniously model 
complex sequences of graph statistics, 
such as degree sequences and shared partner sequences,
without sacrificing interpretability \citep{Hu08,StScBoMo19}. 
A prototypical example used in the social network analysis literature is the geometrically-weighted 
edgewise shared partner model,
which models transitivity through the 
shared partner sequence \citep{SnPaRoHa04, Hu08, StScBoMo19}.

\begin{figure}[t]
\centering
\includegraphics[width=.95\textwidth, keepaspectratio]{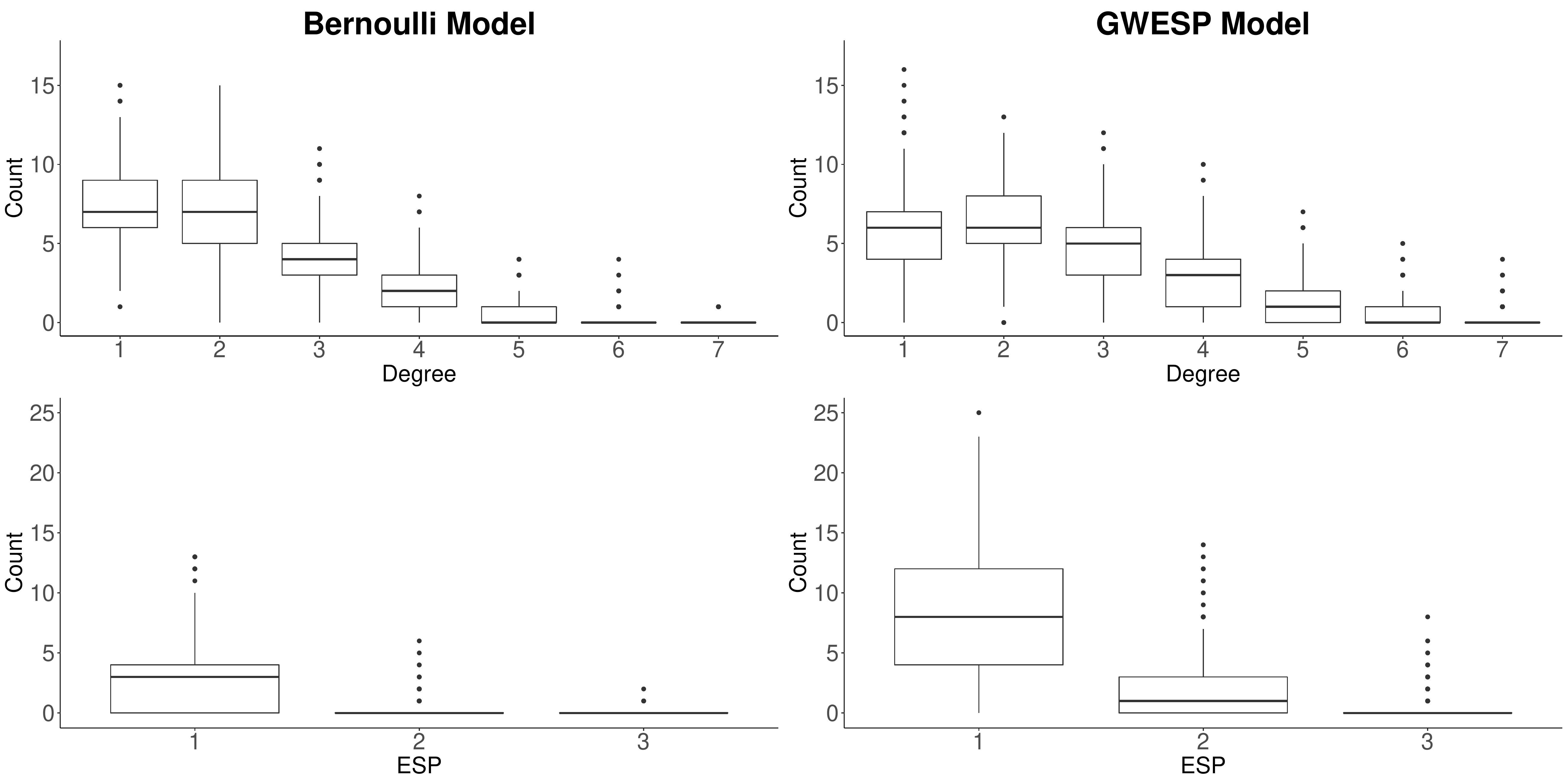}
\caption{\label{fig:gwesp_sim} 
We visualize the degree and ESP distributions of the Bernoulli 
and GWESP model by simulating $1000$ networks from \eqref{eq:curved} 
 with data-generating parameter vector
$(\theta_1, \theta_2, \theta_3) = (-2.5, 0, 1)$ (Bernoulli) and 
$(\theta_1, \theta_2, \theta_3) = (-2.5, .3, 1)$ (GWESP). 
Each column corresponds to each model and we evidence the rightward shift in the degree and ESP distribution of the GWESP model, 
relative to the Bernoulli model. 
}
\end{figure}

We simulate networks according to the following model: 
\be
\label{eq:curved}
\mbP(\bX = \bx)
&\propto& \exp\left(\theta_1 \, \dsum_{i<j}^{N} \, 
x_{i,j} \, + \dsum_{t=1}^{N-2} \, \dsum_{i<j}^{N} \, \eta_t(\theta_2, \theta_3) \, \text{SP}_t(\bx) \right),
\ee
where $\theta_1 \in \mbR$ controls the baseline propensity for edge formation, 
and 
\beno
\eta_t(\theta_2, \theta_3)
\= \theta_2 \, \exp(\theta_3) \, \left[ 1 - (1 - \exp(-\theta_3 \, t) \right],
&& t \in \{1, \ldots, N-2\},  
\ee
parameterizes the sequence of shared partner statistics
\beno
\text{SP}_t(\bx)
\= \dsum_{i<j}^{N} \, x_{i,j} \, \one\left( \, \dsum_{h \neq i,j}^{N} \, x_{i,h} \, x_{h,j} \,=\, t \right),
&& t \in \{1, \ldots, N-2\}. 
\ee
In words,
$\text{SP}_t(\bx)$ counts the number of edges in the network between nodes which have exactly $t$ mutual connections,
commonly called shared partners in the social network analysis literature. 
While $\theta_2 \in \mbR$, in typical applications 
$\theta_2 \geq 0$ and $\theta_3 \in (0, \infty)$,
as values of $\theta_3 < - \log \, 2$ correspond to models which are unstable in the sense of \citet{Sc09b},
and empirical evidence suggests that $\theta_3 \in (0, \infty)$ in many applications 
\citep{Sc09b,StScBoMo19}. 
The effect that the GWESP model specified by \eqref{eq:curved} has on the degree and shared partner distributions of networks 
is visualized in Figure \ref{fig:gwesp_sim},
where positive values of $\theta_2$ stochastically encourage network formations with more transitive edges, 
i.e.,
edges between nodes with at least one shared partner, 
relative to the Bernoulli random graph model with $\theta_2 = 0$.  This is evidenced by the rightward shift in the ESP distribution of the GWESP model, 
relative to the Bernoulli model.

\begin{figure}[t]
\centering
\includegraphics[width=.48\textwidth, keepaspectratio]{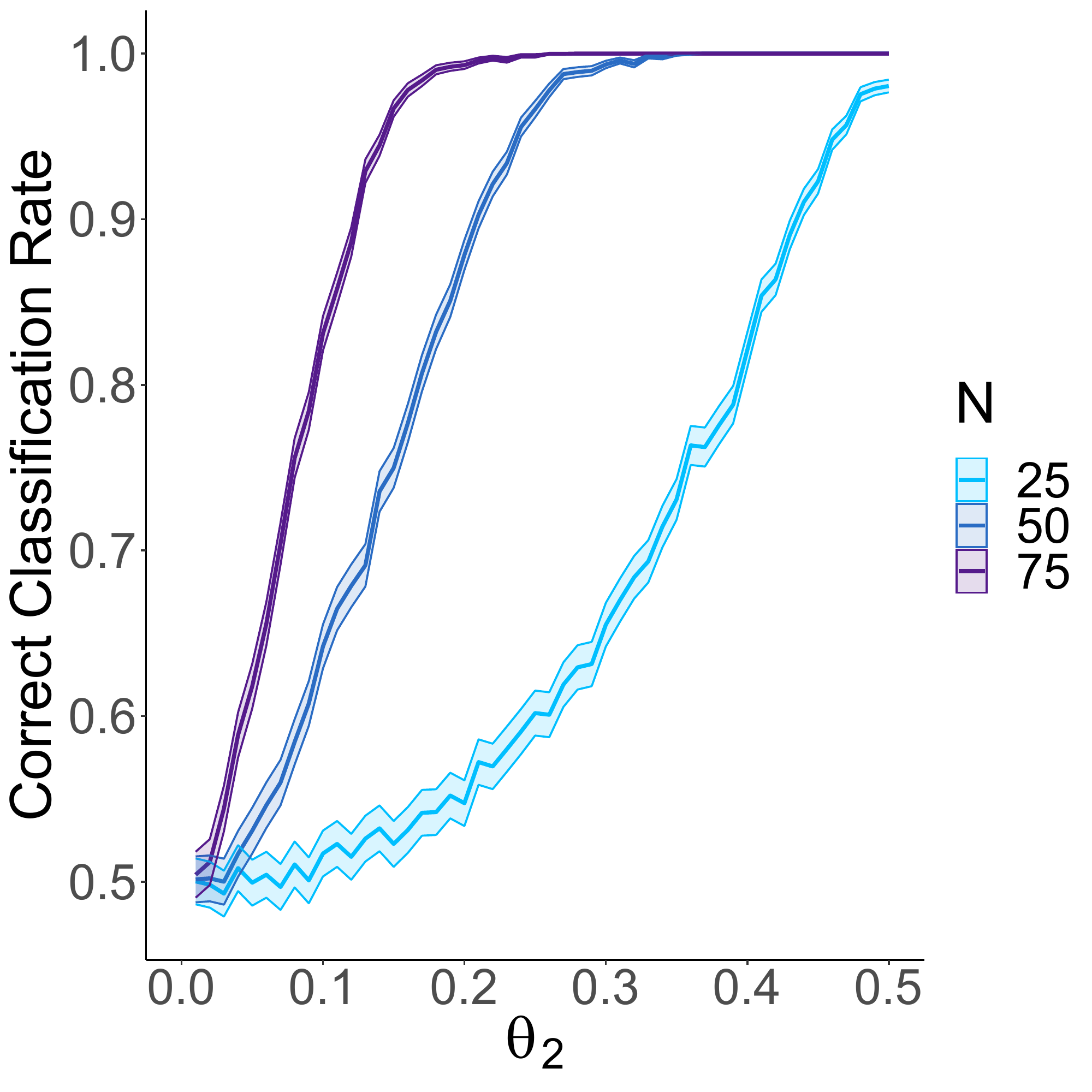} %
\includegraphics[width=.48\textwidth, keepaspectratio]{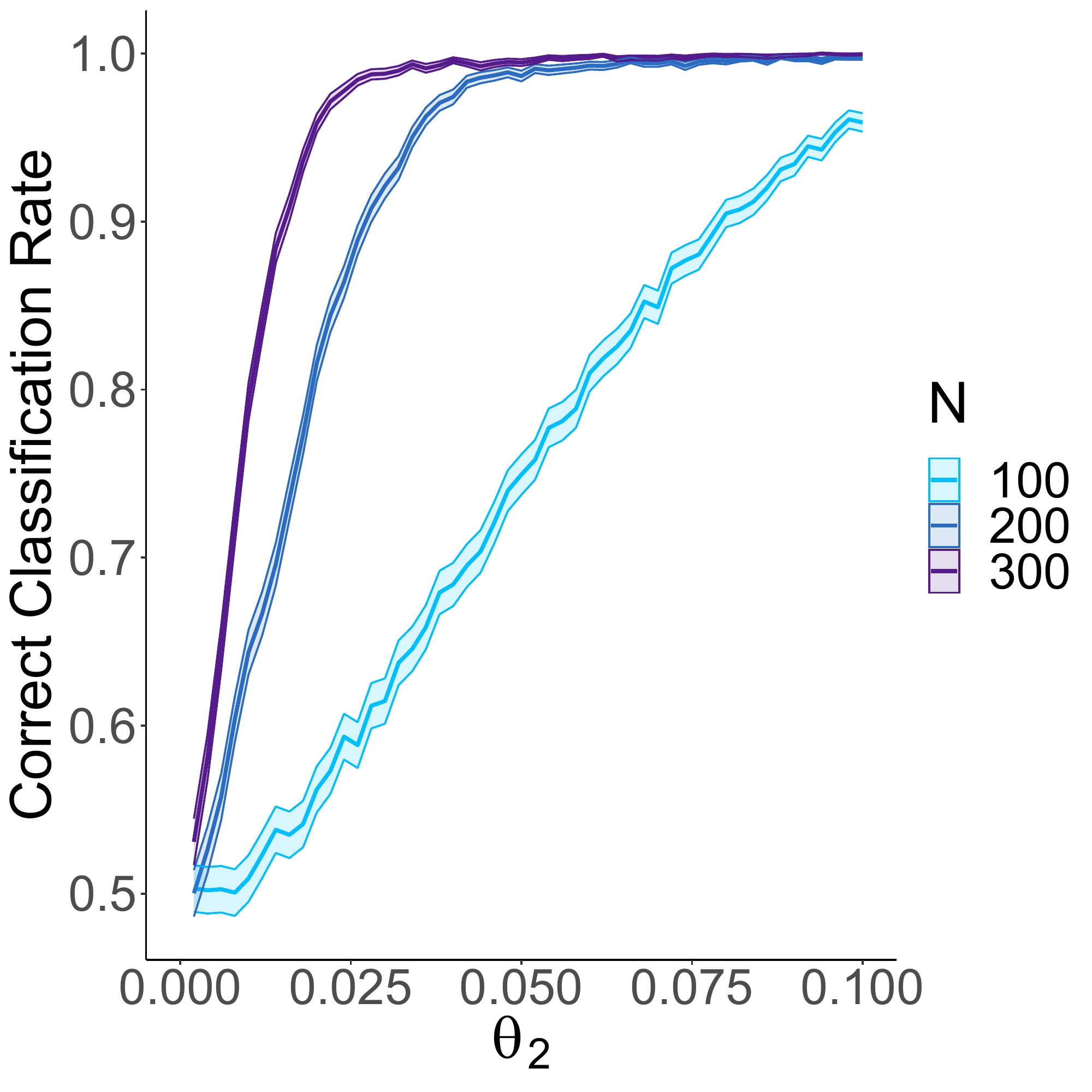} 
\caption{\label{fig:gwesp_res} Results of Simulation study 1. 
(left) Estimate of the correct classification rate 
with $95\%$ confidence band for networks of sizes  
$N = 25, 50, 75$.
(right) Estimate of the correct classification rate with $95\%$ confidence band for networks of sizes 
$N = 100, 200, 300$.}
\end{figure}

We take the true data-generating model $\mM^\star$ to be the curved exponential family specified by  
\eqref{eq:curved}
with parameter vector $\btheta^\star = (-2.5, \theta_2, 1)$,
with $\theta_2$ on a grid covering the interval $[0, \, 0.5]$. 
Note that when $\theta_2 = 0$,
the model reduces to a Bernoulli random graph model with edge probability $p = [1+\exp(-2.5)]^{-1}$. 
We consider the problem of selecting between two models $\mM_1$ and $\mM_2$,
where $\mM^\star = \mM_1$ and $\mM_2$ is the Bernoulli random graph model with edge probability $p = [1+\exp(2.5)]^{-1}$. 
By varying $\theta_2$ we are able to study the threshold of effect size $(\theta_2)$
for which we are able to correctly detect the presence of transitivity in the network, 
as modeled by the  geometrically-weighted edgewise shared partner model in \eqref{eq:curved}.

We vary the network size $N = 25, 50, 75, 100, 200, 300$, 
performing $5000$ replicates for each network size. 
The results of this simulation study  are summarized in Figure \ref{fig:gwesp_res}. 
When $\theta_2$ is close to $0$, 
the point at which $\mM_1 = \mM_2$,
as discussed above, 
our methodology tends to select $\mM_1$ and $\mM_2$ with equal probability.  
However, 
once $\theta_2$ is sufficiently large (relative to the network size $N$), 
our methodology correctly selects $\mM_1$ in almost every replicate.
The effect of the size of the network is seen as we vary 
$N$ from $25$ to $300$.  
When the network size is larger ($N = 100, 200, 300$),
we are able to correctly find the data-generating model $\mM_1$ with high probability  
for smaller values of $\theta_2$. 
In contrast, 
we require $\theta_2 \geq .25$ before we are able to have a high confidence in correctly selecting the data-generating model 
in networks of size $N = 75$,
requiring $\theta_2 \geq .5$ for networks of size $N = 25$.

\subsection{Simulation study 2: reciprocity in directed networks} 

When the adjacency matrix $\bX$ is undirected, 
the corresponding Laplacian matrix $\bL(\bX)$ will be positive semidefinite \citep{BaHw12}, 
resulting in a real-valued vector of eigenvalues $\blambda \in \mbR^N$. 
However, 
when $\bX$ is the adjacency matrix of a directed network, 
the graph Laplacian, as defined for undirected networks, may not be positive semidefinite,
and may involve complex valued eigenvalues.  A common adaptation for directed networks in the literature is to consider the incidence matrix 
$\bB \in  \{0, 1,-1\}^{N \times |E|}$, where $|E|$ is the total number of edges in the network. On each column of the incidence matrix exactly one element will be $-1$, indicating the node where an edge begins, and exactly one element will be $1$, indicating the node where said edge ends. Every other entry is zero. In this manner, a directed network is completely specified by listing all existing edges as columns that indicate which nodes are connected and an orientation between them. We can adapt our proposed methodology to directed networks by considering the symmetric graph Laplacian defined by $\textbf{L}:= \textbf{B}^t\textbf{B}$ \citep{BaHw12}. 

\begin{figure}[t!]
\centering
\includegraphics[scale= 0.4, keepaspectratio]{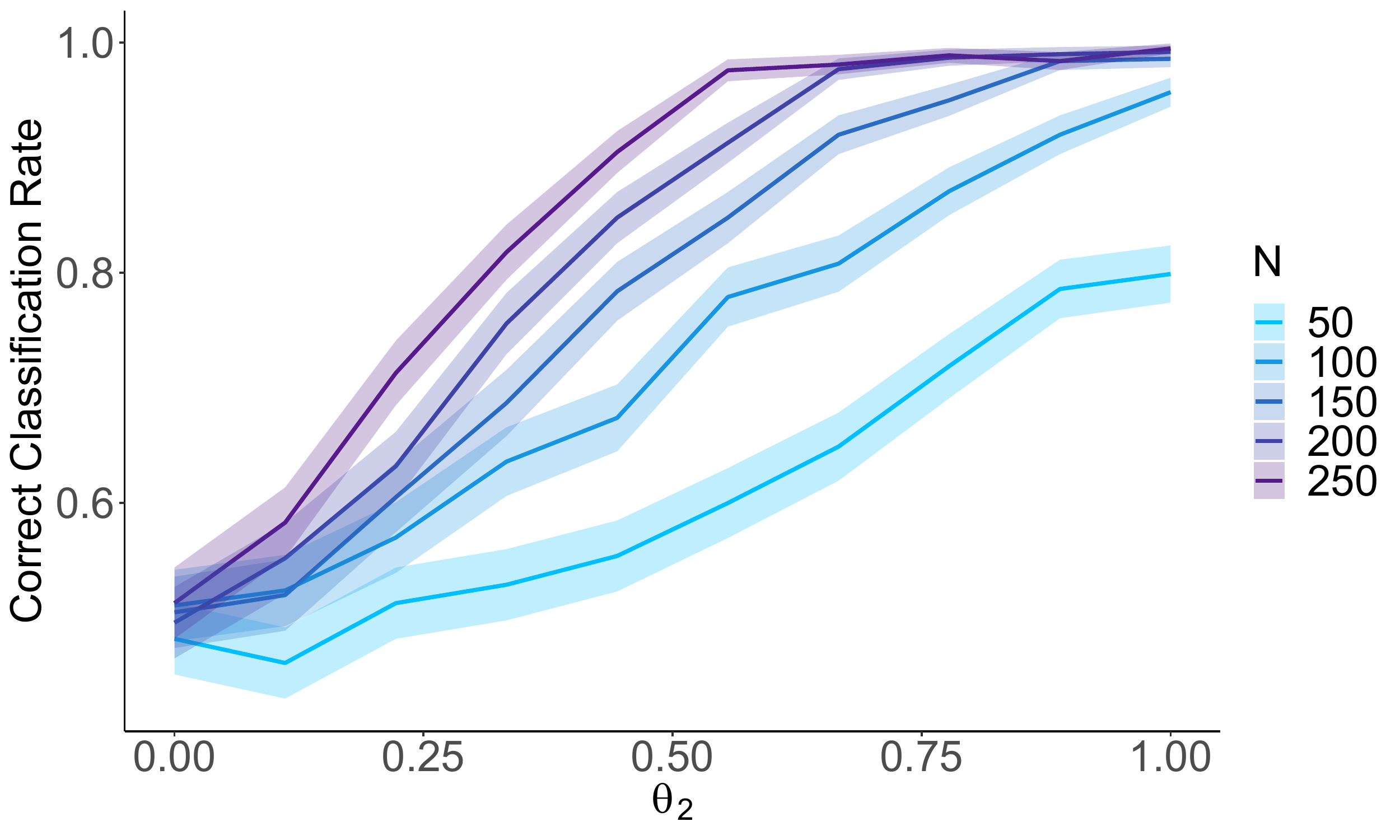} 
\caption{\label{fig:mutual_proportion} 
Results of Simulation study 2. 
Estimates of the correct classification rate with $95\%$ confidence band for various network sizes $N$.
}
\end{figure}

We simulate directed networks from the probability mass function  
\be
\label{eq:directed}
\mbP(\bX = \bx) 
&\propto& \dprod_{i < j}^{N} \, \exp\left( \theta_1 \, (x_{i,j} + x_{j,i})  + \dfrac{\theta_2}{2} \, x_{i,j} \, x_{j,i} \right),
\ee 
We apply our methodology taking $\mM_1$ to be the density only model with fixed $\theta_2 = 0$ in \eqref{eq:directed}. We take
$\mM^\star = \mM_2$ to be the general model specified via \eqref{eq:directed} with unrestricted parameters. 
We conduct a simulation study by taking $\theta_1 = -2.5$ in both $\mM_1$ and $\mM_2$,
taking $\theta_2 = 0$ in $\mM_1$,
and varying $\theta_2$ on a uniform grid of 100 values in $[0,1]$ for $\mM_2$.
The simulation results in Figure \ref{fig:mutual_proportion} are based on $1000$ replications in each case,
reconfirming findings in the previous simulation study which suggested that 
the ability of our methodology to detect the true data-generating model depends on how far $\theta_2$
is from $0$, 
the point at which $\mM_1 = \mM_2$, 
and the size of the network. 
Moreover, 
this study uniquely demonstrates that our methodology can be applied successfully to directed networks.  

\hide{
showing that as $|\theta_2|$ moves farther from $0$ (effect size)  
We simulated $M = 100$ networks from each model and built a design matrix from the eigenvalues of the symmetric Laplacian, defined above. We attempted to classify an observed network from model $M_2$ for different sample sizes $N= 50,100,150,200$, and registered the classification rate after $R=1000$ replications. Figure \ref{fig:mutual_proportion} presents the results of our methodology applied to the symmetric Laplacian for a directed network. We can see how, as the models grow apart or the networsize increases, our methodology correctly identifies the right model class to the observed network.
}

\subsection{Simulation study 3: latent position models} 

Latent variable models for networks,
especially latent position models, 
have witnessed increased popularity and attention since the seminal 
work of \citet{HoRaHa02}.
In this class of models,
nodes are given a latent position $\bz_i \in \bZ$ ($i = 1, \ldots, N$) in a latent space, 
typically taken to be the Euclidean space (i.e., $\bZ = \mbR^k$),
although alternative spaces and geometries have been proposed as well, as is the case of ultrametric spaces \citep{ScSn03}, 
dot product similarity resulting in bilinear forms \citep{HoRaHa02,dot-product-graph}, 
as well as hyperbolic \citep{Krioukov10} and elliptic geometries \citep{SmAsCa19}.
Edges in the network are assumed to be conditionally independent given the latent positions of nodes.
Following \citet{HoRaHa02}, 
we simulate networks in this study accordingly:  
\be
\label{eq:latent}
\log \, \dfrac{\mbP(X_{i,j} = 1 \,|\, \bz_i, \bz_j)}{\mbP(X_{i,j} = 0 \,|\, \bz_i, \bz_j)}
\= \theta - \norm{\bz_i - \bz_j}_2,  
\ee 
where $\theta \in \mbR$ and 
$\bz_i, \bz_j \in \mbR^k$.  
Under this specification, 
the odds of two nodes forming an edge decreases in the Euclidean distance 
$\norm{\bz_i - \bz_j}_2$
between the positions of the two nodes in the latent metric space.

\begin{figure}[t!]
\centering
\includegraphics[width=.98\textwidth, keepaspectratio]{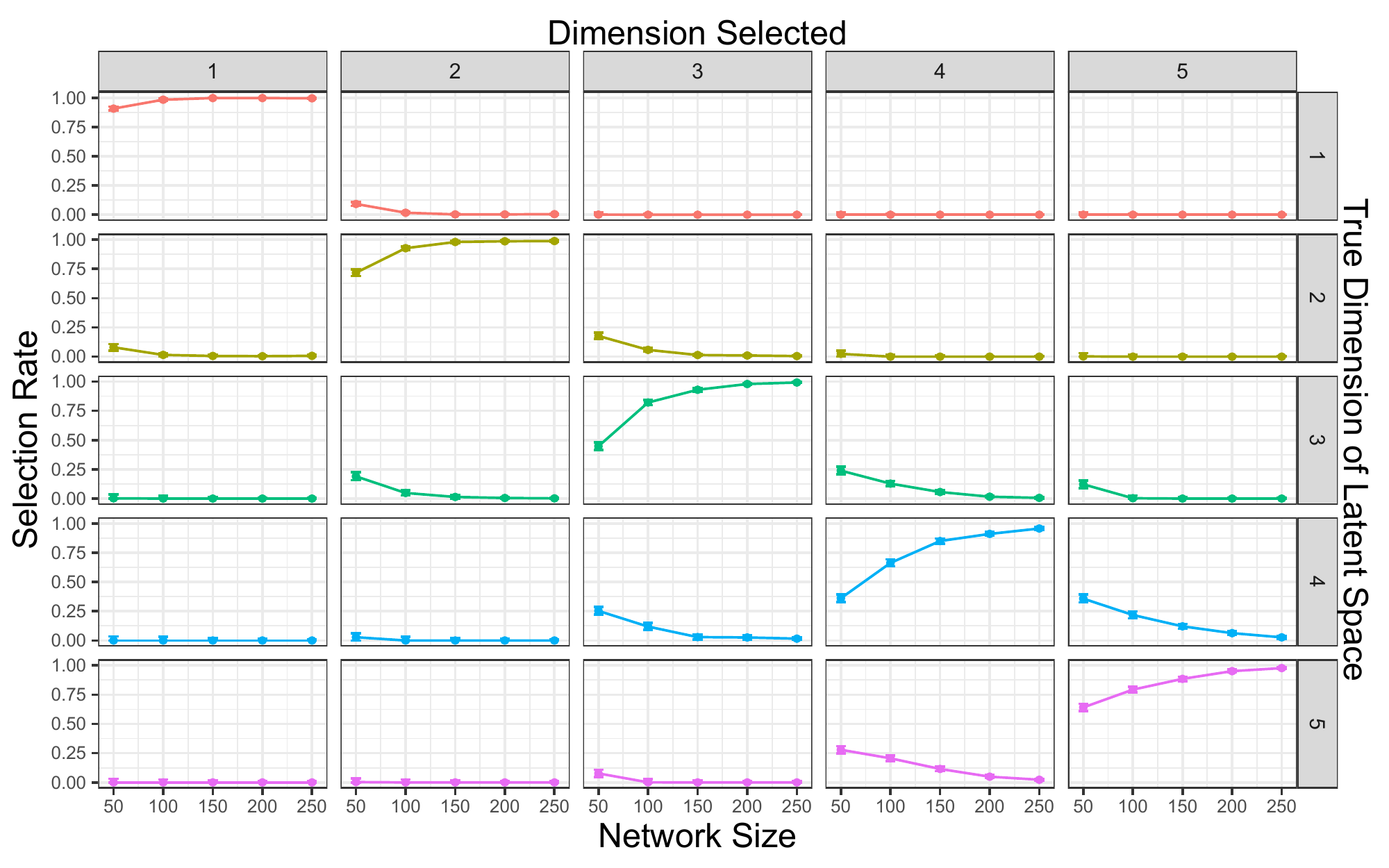} %
\caption{\label{fig:latent} Results of Simulation study 3. 
Estimates of the correct classification rate with $95\%$ confidence intervals for selected network sizes 
and across various latent space dimensions. 
The diagonal panels correspond to correct classification where the selection rate is desired to be highest.  
}
\end{figure}

We explore the ability of our methodology to detect the true dimension of a latent space
by generating networks from the latent Euclidean model described above,
varying the dimension of the latent metric space $k \in \{1,2,3,4,5\}$.
Latent positions of nodes are randomly generated from a multivariate normal distribution in dimension $k \in \{1,2,3,4,5\}$ with zero mean vector and identity covariance matrix.  The candidate competings models are generated in the same fashion across dimensions $1,\ldots, 5$.
We set $\theta = -2.5$ to ensure a low baseline probability of edge formation,
reflecting the sparsity of many real-world networks,
and vary the network size $N \in \{50, 100, 150, 200, 250\}$.  
We apply our model selection methodology in each case and 
compute the percentage of times our methodology selects each of the candidate latent space models.

We summarize the results of the simulation study in Figure \ref{fig:latent}, 
which demonstrates that our methodology is able to correctly identify the true dimension of the 
data-generating latent space model provided the network size is sufficiently large. 
The diagonal panels in Figure \ref{fig:latent} correspond to correct selection of the dimension of the latent space.
Of particular note, the problem becomes more challenging as the dimension of the latent space grows,
but this effect is mitigated as the network size increases, 
with most correct selection rates in this study close to $1$ for networks of size $N = 250$.

\subsection{Simulation study 4: comparing different latent mechanisms} 

\begin{figure}[t!]
\centering
\includegraphics[width=.98\textwidth, keepaspectratio]{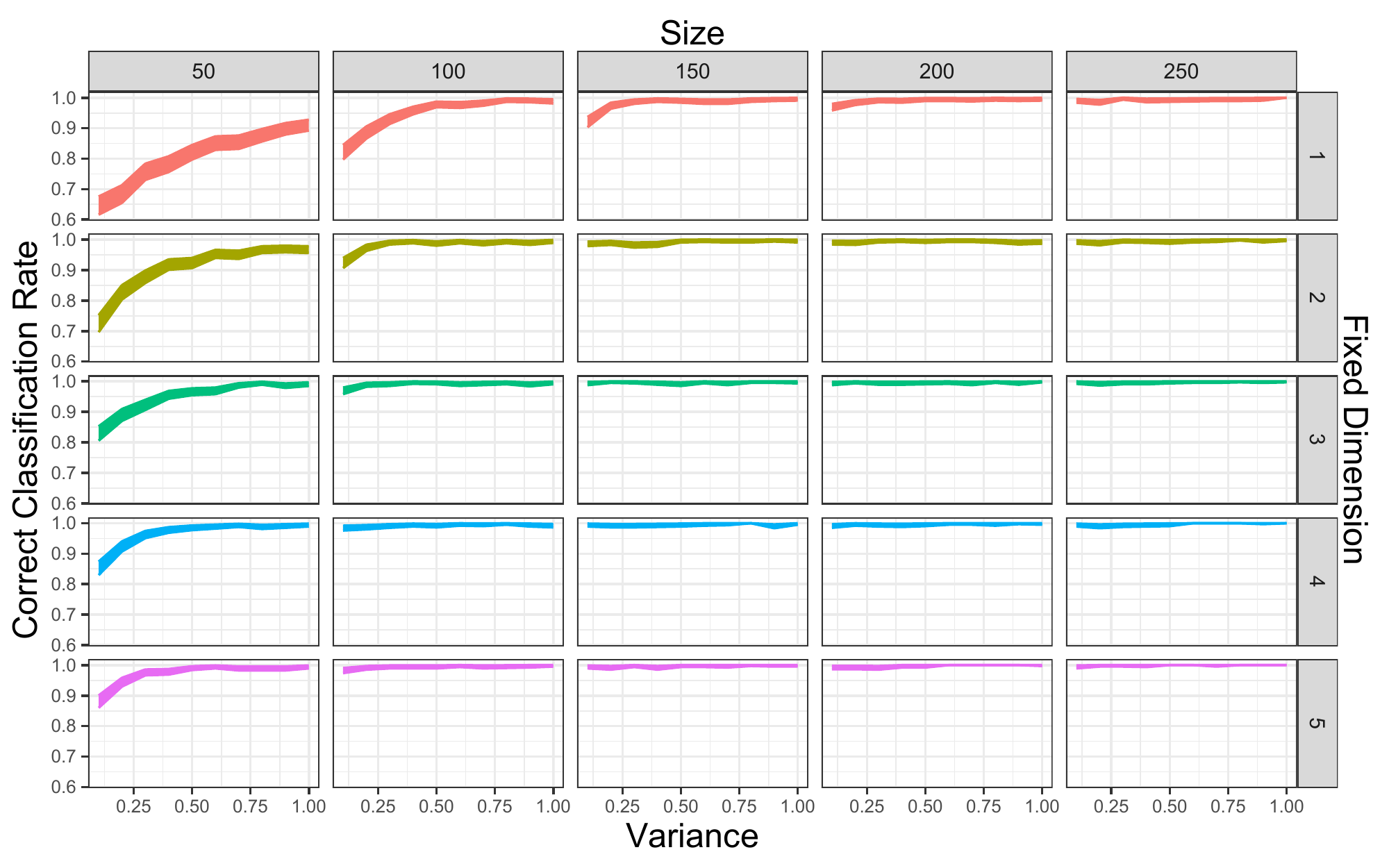} %
\caption{\label{fig:latent_space_eu} Results of Simulation study 4 comparing a distance based model (true model)
to a similarity based model.
Estimates of the correct classification rate with $95\%$ confidence band for different networks 
at different sizes and across different dimensions of latent spaces. 
}
\end{figure}

We next study whether our proposed methodology is capable of distinguishing 
different latent mechanisms for edge formation in a latent position model. 
The first one is the same latent space model 
specified in \eqref{eq:latent}, while the second one replaces the Euclidean distance term 
$-\norm{\bz_i - \bz_j}_2$ with 
the dot product $\bz_i^{t} \bz_k$, 
commonly referred to as a bilinear form. 
A related class of latent position models which utilize bilinear forms of latent node positions 
are random dot product graphs  \citep{dot-product-graph}. 
As in the previous simulation study, 
latent positions of nodes are randomly generated from a multivariate normal distribution with zero mean vector
but this time with covariance matrix $\sigma^2 \, \bI$,
with $\bI$ being the identity matrix (of appropriate dimension) and $\sigma^2 \in \{0.1, 0.2, \ldots, 1.0\}$ a scale factor. 
As the scale factor tends to zero, both models converge to a density only model so detecting the true generating process becomes more difficult.
We summarize the results of the simulation study in Figure \ref{fig:latent_space_eu},
which demonstrates that our methodology is able to correctly identify the true model (distance based) when compared to a bilinear (similarity based) model. 
Of particular note, 
performance improves as the dimension of the latent space increases
and as the network size increases,
as in the previous studies conducted.

\subsection{Simulation study 5: effect of the choice of classifier}

\begin{figure}[t!]
\centering
\includegraphics[width = \linewidth, keepaspectratio]{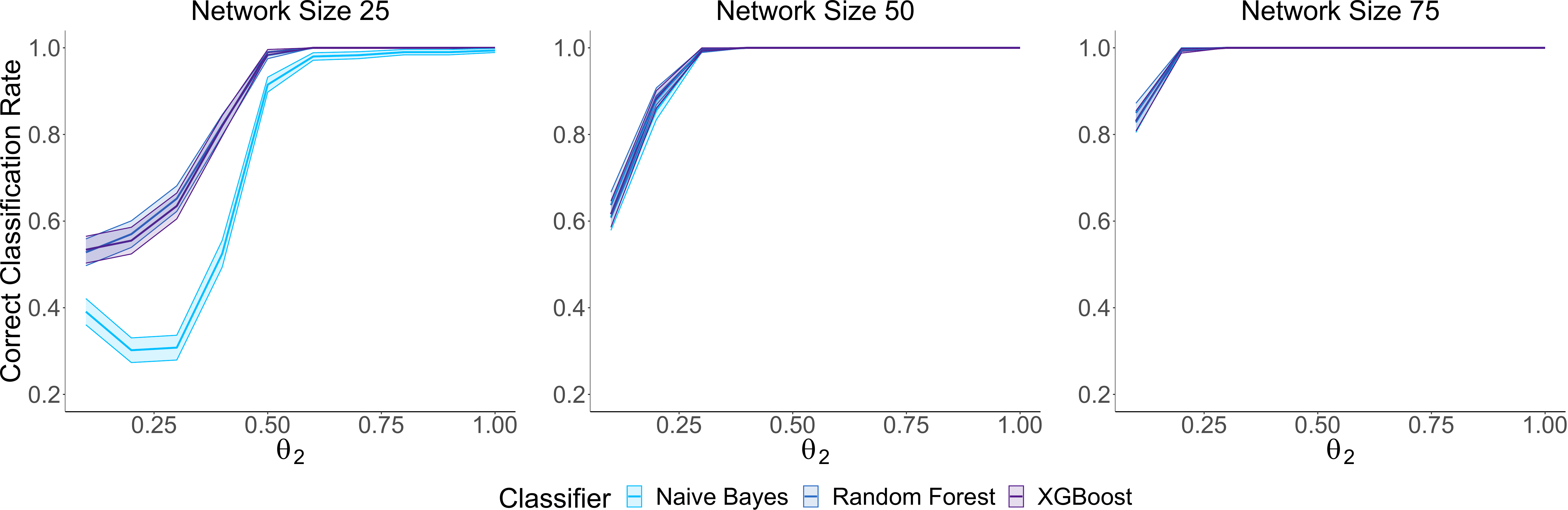} %
\caption{\label{fig:multi} Results of Simulation study 5. 
Estimates of the correct classification rate with $95\%$ confidence band for different classifiers. 
}
\end{figure}

In this study, 
we repeat Simulation study 1 using three different classifiers, 
XGBoost \citep{Chen16}, Random Forest \citep{Ho95, LiWei02} and Naive Bayes \citep{HaYu01,Ma19}. 
Doing so allows us to examine the effect that the choice of classifier has on the results of this simulation study, 
as well as to explore the relative effectiveness of each classifier in this simulation study. 
Figure \ref{fig:multi} shows a similar performance for all classifiers in this simulation study,
with the notable exception being the naive Bayes classifier when networks are size $25$,
suggesting that the choice of classifier has a weak effect on the performance of our proposed methodology, 
provided the network is sufficiently large. 
In line with conclusions in the previous simulation studies,
larger network sizes result  
in more pronounced model signatures.
In light of these results, 
the effect of the choice of classifier appears to diminish if the model signal is sufficiently strong.

\section{Applications} 
\label{sec:applications}  

In order to study the performance of our proposed model selection methodology in applications to real-world network data, 
we study two network data sets which have previously been studied in the literature, in order to have a baseline for evaluating whether our methodology confirms existing results and knowledge about these networks. 

\subsection{Application 1: Sampson's monastery network} 

We apply our model selection methodology to the Sampson's monastery network data on social relationships (likeness) among 18 monk novices in a New England monastery in 1968 \citep{Sa68}. 
Based on the existing literature studying this network, 
we propose different model structures for this network which are well-designed to capture the 
community structure known to be a critical component of the network. 
In order to model this structure, 
stochastic block models have been applied to the network \citep{ABFX08},
as well as latent position models with a hierarchical group-based prior distribution structure on the latent positions 
\citep{HaRaTa07}. 
We consider the following models: 
\bi
\item SBM: $\mM_1$--$\mM_4$ correspond to stochastic block models with $K = 1, 2, 3, 4$ blocks ($\mM_1$ being equivalent to a density only model). 
\item LPM: $\mM_5$--$\mM_8$ correspond to latent position models with model terms for density and reciprocity and 
latent space dimensions $K = 1, 2, 3, 4$.  
\item GLPM:  $\mM_9$--$\mM_{20}$ combine the two previous specifications by utilizing the hierarchical group-based prior distribution structure of \citet{HaRaTa07}, 
considering all combinations of group number $K = 2, 3, 4$ and latent space dimension $d= 1, 2, 3, 4$. 
\ei

{  
\begin{table}[t!]
\centering 
\begin{tabular}{|c|c|c||c|c|c|}
\hline
\begin{tabular}[c]{@{}c@{}}Model\end{tabular}   & $s_i$ & $\tilde{s}_i$  & \begin{tabular}[c]{@{}c@{}}Model\end{tabular}   & $s_i$  & $\tilde{s}_i$   \\ \hline\hline 
\begin{tabular}[c]{@{}c@{}} $\mM_1$\\ ({\footnotesize SBM}, $\text{\footnotesize $K=1$}$)\end{tabular}      & 0.002      & 0.004       & \begin{tabular}[c]{@{}c@{}} $\mM_2$\\ ({\footnotesize SBM}, $\text{\footnotesize $K=2$}$)\end{tabular}      & 0.003          & 0.007          \\ \hline
\begin{tabular}[c]{@{}c@{}} $\mM_3$ \\ ({\footnotesize SBM}, $\text{\footnotesize $K=3$}$)\end{tabular}      & 0.032      & 0.077       & \begin{tabular}[c]{@{}c@{}} $\mM_4$\\ ({\footnotesize SBM}, $\text{\footnotesize $K=4$}$)\end{tabular}      & \textbf{0.410} & \textbf{1} \\ \hline
\begin{tabular}[c]{@{}c@{}} $\mM_5$\\ ({\footnotesize LPM}, $\text{\footnotesize $d=1$}$)\end{tabular}      & 0.028      & 0.068       & \begin{tabular}[c]{@{}c@{}} $\mM_6$\\ ({\footnotesize LPM}, $\text{\footnotesize $d=2$}$)\end{tabular}      & 0.028          & 0.069          \\ \hline
\begin{tabular}[c]{@{}c@{}} $\mM_7$ \\ ({\footnotesize LPM}, $\text{\footnotesize $d=3$}$)\end{tabular}      & 0.005      & 0.013       & \begin{tabular}[c]{@{}c@{}} $\mM_8$ \\ ({\footnotesize LPM}, $\text{\footnotesize $d=4$}$)\end{tabular}      & 0.003          & 0.008          \\ \hline
\begin{tabular}[c]{@{}c@{}} $\mM_9$ \\ ({\footnotesize GLPM}, $\text{\footnotesize $K=2,d=1$}$)\end{tabular}  & 0.023      & 0.055       & \begin{tabular}[c]{@{}c@{}} $\mM_{10}$ \\ ({\footnotesize GLPM}, $\text{\footnotesize $K=3,d=1$}$)\end{tabular} & 0.044          & 0.108          \\ \hline
\begin{tabular}[c]{@{}c@{}} $\mM_{11}$ \\ ({\footnotesize GLPM}, $\text{\footnotesize $K=4,d=1$}$)\end{tabular} & 0.043      & 0.104       & \begin{tabular}[c]{@{}c@{}} $\mM_{12}$\\ ({\footnotesize GLPM}, $\text{\footnotesize $K=2,d=2$}$)\end{tabular} & 0.060          & 0.147          \\ \hline
\begin{tabular}[c]{@{}c@{}} $\mM_{13}$ \\ ({\footnotesize GLPM}, $\text{\footnotesize $K=3,d=2$}$)\end{tabular} & 0.083      & 0.202       & \begin{tabular}[c]{@{}c@{}} $\mM_{14}$ \\ ({\footnotesize GLPM}, $\text{\footnotesize $K=4,d=2$}$)\end{tabular} & 0.036          & 0.089          \\ \hline
\begin{tabular}[c]{@{}c@{}} $\mM_{15}$ \\ ({\footnotesize GLPM}, $\text{\footnotesize $K=2,d=3$}$)\end{tabular} & 0.020      & 0.050       & \begin{tabular}[c]{@{}c@{}} $\mM_{16}$ \\ ({\footnotesize GLPM}, $\text{\footnotesize $K=3,d=3$}$)\end{tabular} & 0.041          & 0.101          \\ \hline
\begin{tabular}[c]{@{}c@{}} $\mM_{17}$ \\ ({\footnotesize GLPM}, $\text{\footnotesize $K=4,d=3$}$)\end{tabular} & 0.061      & 0.148       & \begin{tabular}[c]{@{}c@{}} $\mM_{18}$\\ ({\footnotesize GLPM}, $\text{\footnotesize $K=2,d=4$}$)\end{tabular} & 0.012          & 0.029          \\ \hline
\begin{tabular}[c]{@{}c@{}} $\mM_{19}$\\ ({\footnotesize GLPM}, $\text{\footnotesize $K=3,d=4$}$)\end{tabular} & 0.030      & 0.074       & \begin{tabular}[c]{@{}c@{}} $\mM_{20}$ \\ ({\footnotesize GLPM}, $\text{\footnotesize $K=4,d=4$}$)\end{tabular} & 0.035          & 0.085          \\ \hline
\end{tabular}
\caption{\label{tab:sampson} Propensity scores $s_i$ and normalized propensity scores 
$\tilde{s}_i$ for models $\mM_1$--$\mM_{20}$ for the Sampson's monastery network. 
}
\end{table}

}

\begin{figure}[t!]
\centering
\includegraphics[scale= 0.4, keepaspectratio]{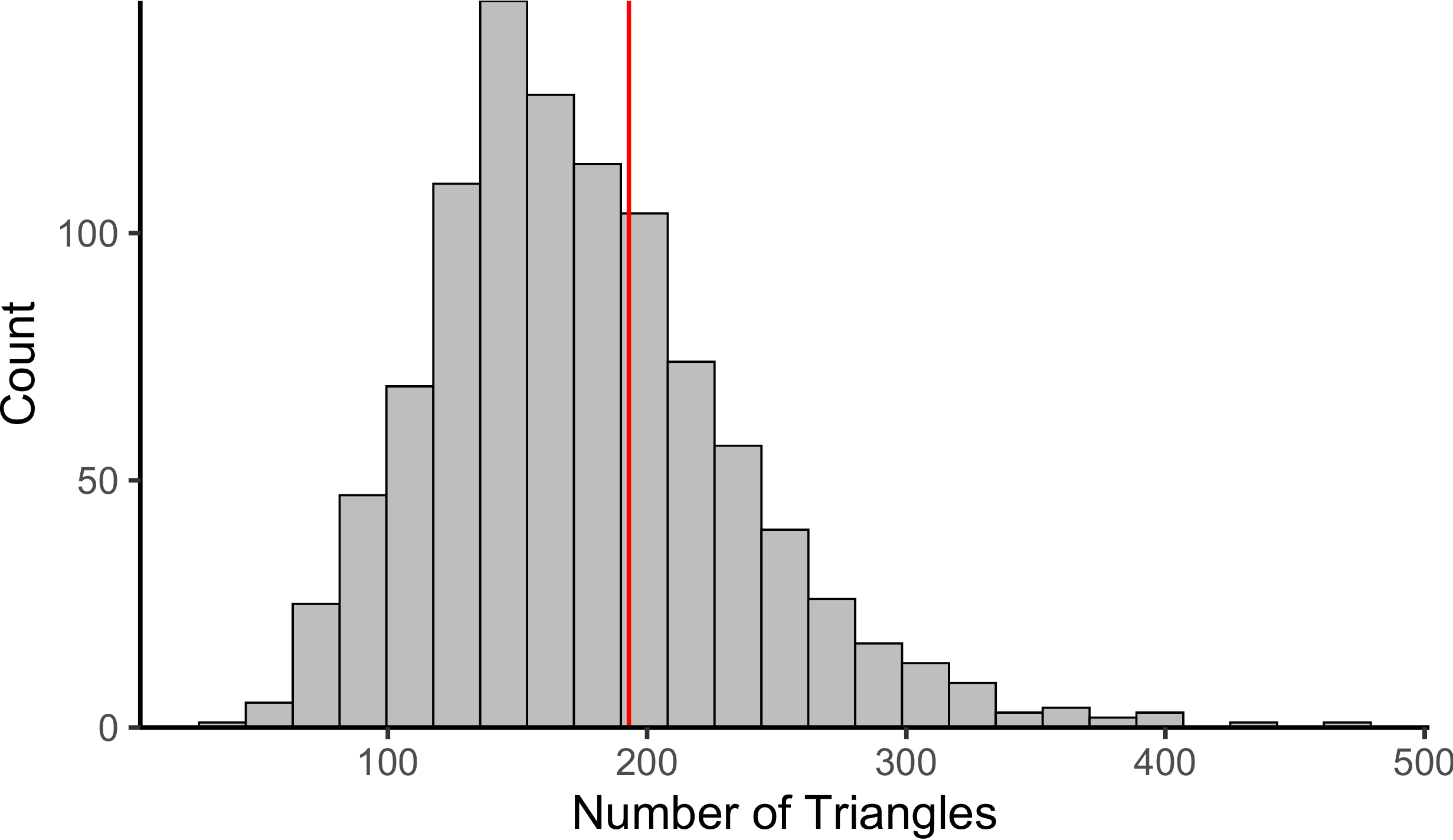} 
\caption{\label{fig:plot_triangles} Fit of the observed number of triangles in the Sampson network relative to the distribution of triangles from simulated networks from  $\mM_4$. The observed number of triangles is indicated in red.}
\end{figure}

Each model was fit and our model selection methodology was applied to choose a best fitting model. 
The latent space models were fit with \citet{latentnet} and the stochastic block models were fit with \citet{Leger16}. 
Table \ref{tab:sampson} presents the results. 
The model with the highest propensity score is $\mM_4$, the stochastic block model with $K = 4$ blocks. 

It has been well-established in the literature that the Sampson's monastery network features strong community structure 
\citep{HaRaTa07,ABFX08},
featuring three labeled groups. 
However,
statistical analyses have revealed the presence of a potential fourth group,
evidenced in analysis which employ mixed membership stochastic block models \citep{ABFX08},
as well as evidence in studies which employ latent position models which suggests 
certain nodes may have strong connections to two or more labeled groups \citep{HaRaTa07}. 
Within the context of the models we considered here, 
the choice of a stochastic block model with $K = 4$ blocks appears to be sufficient to capture 
the mixing patterns of the communities as well as the reciprocity from the inclusion of a reciprocity term. 
We hold the opinion that the expression of transitivity is not sufficiently strong in this network, 
otherwise the latent position model with $K = 4$ groups would potentially serve as a better model, 
as latent position models are able to capture network transitivity through the latent metric space. 
Figure \ref{fig:plot_triangles} supports this claim by simulating networks from $\mM_4$ and comparing the empirical 
triangle count distribution of these simulated networks to the observed number of triangles in the network, 
demonstrating good model fit in this regard.  

\subsection{Application 2: multilevel school network}

We end the section with an application to a multilevel network consisting of 6,607 third grade students over 306 classes 
across 176 primary schools in Poland in the 2010/2011 academic year \citep{maluchnik:suek_sample-2014}. 
Our interest in this data set lies in the fact that it has already been extensively studied in \citet{StScBoMo19},
which provides the closest we can get to a data-generating model. 
The network contains 306 classes, 
but features a significant portion of non-response resulting in a large percentage of missing edge data in the network.
The issues of missing data require careful consideration and are beyond the scope of this work. 
As such, 
we restrict our study in this work to the 44 classes within the multilevel network that did not feature 
any missing edge data.  
The data set employed is a directed network of 906 nodes corresponding to the individual students within the 
44 classes without missing edge data, 
where a directed edge $i \to j$ implies that person $i$ stated they were friends person $j$. 
Part of the data collected included the sex of each student (recorded as male or female). 
This multilevel network data set naturally fits into the local dependence framework of \citet{ScHa13},
for which class based sampling is justified under the local dependence assumption 
\citep[Proposition 2 \& Theorem 2,][]{ScSt19};  
additional details of the data set can be found in \citet{StScBoMo19}.

\begin{table}
\centering 
\begin{tabular}{l|cccc}
Model Term    & $\mM_1$    & $\mM_2$    & $\mM_3$    & $\mM_4$ \\ 
\hline 
Edges   & \checkmark & \checkmark & \checkmark & \checkmark \\  
Mutual  & \checkmark & \checkmark & \checkmark & \checkmark \\
Out-degrees
(1--6) & \checkmark & \checkmark & \checkmark & \checkmark \\
\hline 
Out-degree 
(Female)& \checkmark & \checkmark & \checkmark & \checkmark \\
In-degree
(Female)& \checkmark & \checkmark & \checkmark & \checkmark \\ 
Sex-match& \checkmark & \checkmark & \checkmark & \checkmark \\
\hline  
GWESP 
(decay parameter fixed at $0$)  & & \checkmark & & \\ 
GWESP 
(decay parameter fixed at $.25$) & & & \checkmark & \\ 
GWESP 
(decay parameter estimated) & & & & \checkmark \\ 
\hline   
\end{tabular}
\caption{\label{tab:socnet_models} Descriptions of Models 1--4 found in \citet{StScBoMo19}.}
\end{table}

In this application,
we study whether our proposed methodology for model
selection coincides with published findings for this network by studying 
Models 1--4 published in \citet{StScBoMo19},
which we summarize in Table \ref{tab:socnet_models}. 
The first three model terms (edges, mutual, and out-degree terms) control for structural effects within the network, 
including density, reciprocity, and fitting the degree distribution. 
The next three model terms adjust for different sex-based edge effects and homophily. 
The last three model terms correspond to the geometrically-weighted shared partner (GWESP) term 
specified in \eqref{eq:curved} that was studied in Simulation study 1.
The inclusion of this model term is aimed at  
capturing a stochastic tendency towards network transitivity and triad formations
based on values of the base parameter ($\theta_2$ in \eqref{eq:curved}) and the decay parameter 
($\theta_3$ in \eqref{eq:curved}). 
Model 1 includes no GWESP term, 
whereas Model 2 and Model 3 fix the decay parameter at specific values found in the literature,  
reducing the curved exponential family to a canonical exponential family (see discussions in \citet{Hu08} 
and \citet{StScBoMo19}). 
Model 4 estimates the decay parameter. 

%

\begin{figure}[t!]
\centering
\includegraphics[scale= 0.28, keepaspectratio]{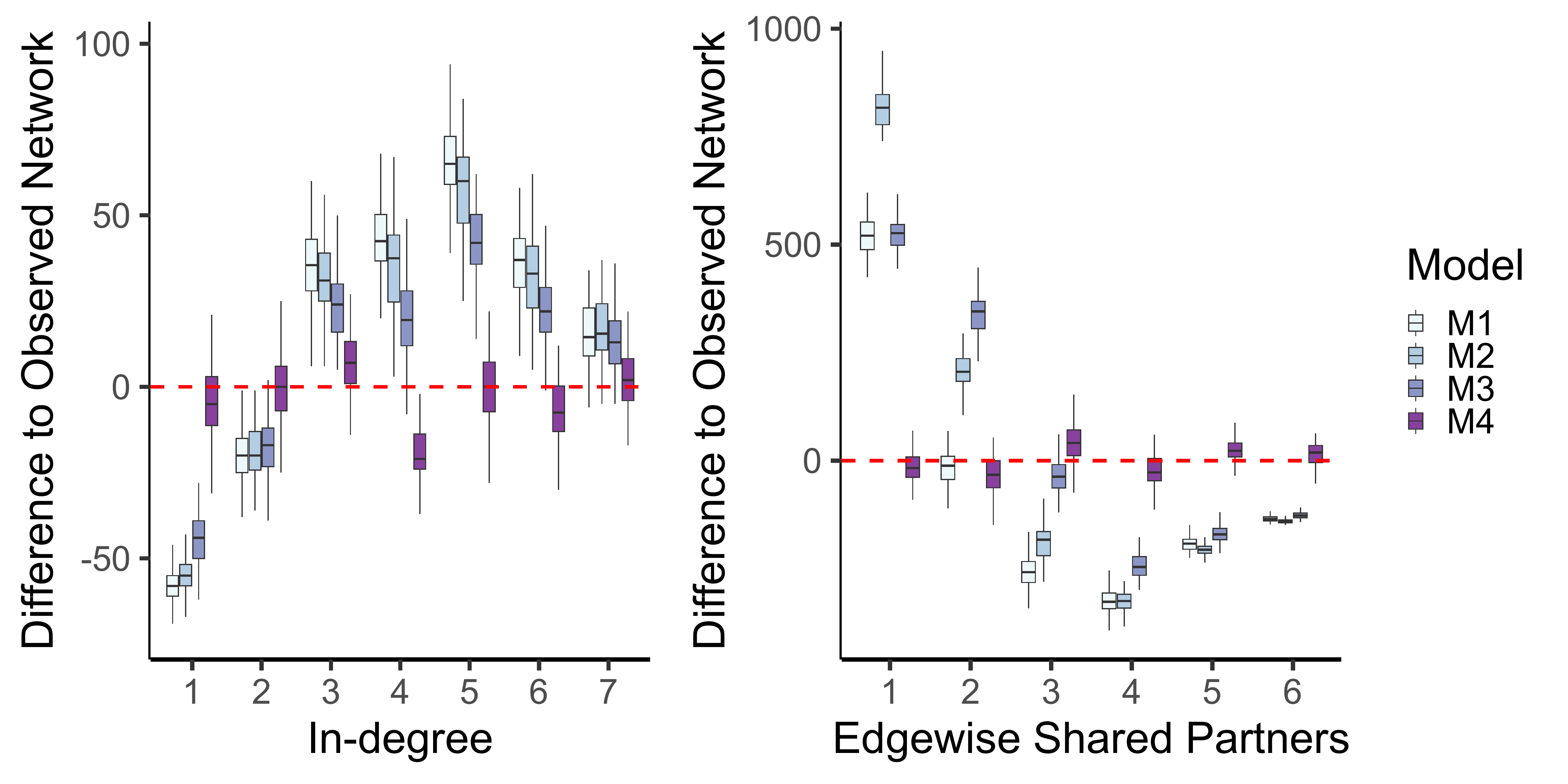} 
\caption{\label{fig:mplot_classes} Difference in selected statistics between fitted models and Polish school network.}
\end{figure}

We fit each of the four models $\mM_1, \mM_2, \mM_3, \mM_4$ 
and apply our model selection methodology, which selects model $\mM_4$ ($\text{propensity}=0.997$) above all other candidate models. This coincides with the findings of \citet{StScBoMo19},
who explored the fit of various models to the data set with respect to common-place heuristic measures 
\citep{HuGoHa08},
as well as out-of-sample measures and through the Bayesian Information Criterion (BIC). 
Figure \ref{fig:mplot_classes} demonstrates the model fit to important network features.  

\hide{
After applying our methodology with $K=100$ simulations, model 4 ($\text{propensity}=0.997$) was preferred over any other fitted model. This is in direct agreement with an analysis on \label{fig:mplot_classes}, that clearly show model 4 does a better job at fitting the observed network in regards to indegree, outdegree and edge-wise shared partners distribution.

}

\bibliographystyle{chicago} 
\bibliography{base.bib}

\end{document}